# Democratic governance and international research collaboration: A longitudinal analysis of the global science network

Travis A. Whetsell[1*]

[1]School of Public Policy, Georgia Institute of Technology, Atlanta, Georgia, United States of America

*Corresponding author
E-mail: travis.whetsell@gatech.edu

# Abstract

The democracy-science relationship has traditionally been examined through philosophical conjecture and single country case studies. There remains limited global scale empirical research on the topic. This study explores country level factors related to the dynamics of the global scientific research collaboration network, focusing on structural associations between democratic governance and the strength of international research collaboration ties. This study combines longitudinal data on 170 countries between 2008 and 2017 from the Varieties of Democracy Institute, World Bank Indicators, Scopus, and Web of Science bibliometric data. Methods of analysis include descriptive network analysis, temporal exponential random graph models (TERGM), and valued exponential random graph models (VERGM). The results suggest positive significant effects of democratic governance on the formation and strength of international research collaboration ties, as well as homophily between countries with similar levels of democratic governance. The results also show the importance of exogenous factors, such as GDP, population size, and geographical distance, as well as endogenous network factors including preferential attachment and transitivity.

# Introduction

Knowledge of the relationship between country-level governance and processes and outcomes of national scientific performance remains limited. Despite a wealth of country level data regarding political, economic, and bibliometric indicators, few studies have examined the association between country level factors and research collaboration or the evolution of international collaboration networks. As such, the relationship between structural factors and collaborative research that spans national borders remains somewhat opaque. Robust research programs that measure indicators of democratic governance [1] [2] as well as indicators of scientific research activity [3] have developed in parallel but remain relatively disconnected.

Addressing questions of governance are important for a full understanding of the antecedents of scientific collaboration. If, as scholars like Karl Popper and Robert Merton suggested, democracy is the optimal system of governance for the flourishing of science, then a large portion of the world's human capital remains untapped. The vast majority of the world's population currently lives under closed or electoral autocracies [4]. Yet, outlier countries such as China have made major advancements in international science in recent decades, despite remaining one of the most autocratic countries in the international system [5, 6]. Further, if democracy is optimal for science, then the recent general trend of democratic backsliding and rising authoritarianism makes the question more salient [4]. Global democratization peaked in 2012 and has since reverted to 1989 levels, erasing the last 30 years of progress. Further, scientists and scholars are increasingly subject to repression by "autocratizing" nations and academic freedom is declining globally [7]. A philosophical assumption of this study is that international collaboration in science is not adequately explained by reference only to facts about individuals or institutions, but rather that a nation's overarching governance structures enable, constrain, and influence the self-organization among scientists and scholars [8, 9].

The current study builds on research concerned with the relationship between country level governance factors and innovation in science and technology [10, 11]. The study tests hypothesized effects of democratic governance on the probability of forming strengthening research ties in the global network of international science. While there are numerous studies spanning disciplines that analyze bibliometric indicators of scientific research, from inputs such as inclusion and inequality [12] to outcomes such as citation impact [13], the current study leverages bibliometric data to construct and analyze a global network of international research collaboration ties and its relationship to liberal democracy. This article also builds on research that examines cross-governance effects on international cooperation [14, 15] to better understand assortative mixing (homophily) between democracies and autocracies. Finally, this article examines more complex processes inherent in a wide array of networks, such as preferential attachment and transitivity, which are important processes that manifest in wide array of social networks [16-18].

Inferential network analysis techniques are applied to test the hypotheses while accounting for the dyadic dependency inherent in network tie formation [19]. These tools provide novel techniques for analyzing international collaboration growth suitable to the relational structure of the data. While previous studies at the international level have used individual level (country level) indicators, this study leverages more recent developments in network science to account for endogenous processes inherent in relational co-authorship data. To this end, valued and temporal exponential random graph models are used to analyze the formation and strength of ties between 170 countries in the international collaboration network. Hypotheses are tested using longitudinal data from the Varieties of Democracy Institute, World Bank Indicators, and Scopus and Web of Science bibliometric data from 2008 to 2017.

# Literature Review and Hypothesis Development

The literature review derives hypotheses by synthesizing inter-disciplinary insights from disparate literatures across scientometric studies of international research collaboration, philosophical scholarship on the nature of science, and political science studies of democracy. Given the lack of extant theory on the direct topic, it is necessary to review these literatures to connect the dots regarding the relationship between national governance structures and international collaboration patterns in scientific research.

## International Research Collaboration

The topic of research collaboration as a subject of inquiry in science policy is wide ranging and well developed. The term "invisible college" captures the early roots of scientific collaboration, referring to a group of seventeenth century scientists who organized activities informally through the collaborative exchange of letters, which later formed into the Royal Society of London. Derek de Solla Price [20] documented a steady acceleration of coauthored papers in the early 20th century, arguing that science had moved from "little science" in the seventeenth century to a contemporary form of "big science" involving national expenditures, economic impacts, computers, rockets, and the Manhattan Project. One of the most astonishing examples of contemporary international research collaboration includes the Atlas Experiment in particle physics, which produced a series of publications with thousands of authors from dozens of countries [21]. In the contemporary setting, the "new invisible college" refers to a sprawling global network of authors connected by papers rather than institutions [8, 22, 23]. The growth and density of the international collaboration network continues to increase [24]. However, relatively few scholars have questioned whether and how the overarching governance structures of nations influence self-organization among scientists in the international system.

Numerous factors are implicated in the dramatic evolution of research collaboration as a mode of scientific inquiry over the past century. The costs of cooperation have declined substantially, particularly given the rise of the internet, email, and video conferencing [25, 26]. The steadily increasing specialization of scientific disciplines [27], as well as the evolution of multidisciplinary research projects [28] have also contributed. The complexity, scope, and magnitude of scientific programs has increased, requiring participation by numerous researchers to execute single experiments, particularly those involving complex and expensive equipment [29]. Further, the institutional norms of higher education institutions, as well as governmental organizations, have adjusted to accommodate and promote collaborative teams [30]. Meanwhile, scholars themselves have adjusted to the new norm, realizing gains in "scientific human capital" and productivity through collaboration domestically and across national borders [31, 32]. The journey toward promotion and tenure through the traditional route of "publish or perish" may be less perilous with collaborative partners shouldering a part of the burden of knowledge creation. Lastly, collaboration facilitates the friendship, career advancement, learning [33], and epistemic goals of scientists [27].

Abramo et al. [34] showed that the association between research collaboration and the scientific productivity of universities varies substantially across disciplines, between basic and applied science, and in the context of domestic versus international collaboration. Recently, Fox and Nikivincze [35] showed that collaboration is an important predictor in prolific publication productivity in science. International research collaboration, as distinct from domestic collaboration, may have even greater returns. As Glanzel and de Lange [36] showed, internationally coauthored papers tend to have higher citation impact than domestically published papers. Reversing the question, Abramo et al. [34] tests whether productivity of the scientist predicts higher level of international research collaboration. Their findings indicate more productive scientists tend to be internationally collaborative. Thus, there appears to be an endogeneity/chicken-egg problem at work, where collaboration begets performance begets collaboration. However, the literature has repeatedly shown that international research collaboration results in higher productivity and performance [37]. Wagner et al. [23] questioned whether international collaboration tended to produce conventional, novel, or atypical research, finding, contrary to expectation, that conventional research tended to be the most typical outcome. Using a different measure of internationalization in research, Sugimoto et al. [38] and find that mobile scientists with many country affiliations in their publication record tend to build up international collaboration networks as they travel within the international system and also tend to have the highest citation impact[39]. Similarly, Robinson-Garcia et al. [40] found that mobile scientists tend to have higher publication output and citation impact.

    Fewer studies have examined the effects of national level characteristics. Glanzel and Schubert [41] suggest that larger countries, such as India, Japan, United States, and Turkey tend to produce a higher proportion of domestic publications than smaller countries. Conversely, smaller or more remote countries, such as Switzerland, Austria, South Korea, tend to have higher international rates of publication. Leydesdorff et al. [42] recently examined whether the influence

of government funding or international research collaboration effects a nation's citation impact finding the balance favors international research collaboration.

Finally, Subramanyam [43] suggested coauthorship as an indicator of collaboration has its advantages as a measure, being non-reactive, invariant, quantifiable, and inexpensive. However, Katz and Martin [29] cautioned against reading too much into the term 'collaboration'. As Davidson and Carpenter [44] suggest there are also ambiguities that should be appreciated in international collaboration, such as single authorship with multiple affiliations, but also suggested such errors are small relative to the size of bibliometric databases. Thus, international co-authorship may not always reflect true collaboration. Rather, the approach of this study is that the term "international research collaboration" refers minimally to instances in which two or more affiliations referring to distinct countries co-occur on a single research paper.

## Governance Effects on Innovation in Science and Technology

The rise of China in the international system of science increases the salience of questions regarding democratic governance and scientific capacity. These are traditionally assumed to be complementary if not critically compatible. But few large-scale empirical studies have examined the general effect of governance on scientific processes and outcomes, or on how the relationship might influence international research collaboration. Scholars suggest that social structure influences organizations and individuals, who themselves reproduce and sometimes transform the structures within which they are embedded [9, 45]. However, while a large body of research has focused on the individual, the discipline, and the institution, the effects of political governance structures on the self-organization of scientific praxis is not empirically well established.

The conjecture that democracy is the optimal political environment for the advancement of science has existed since both popular sovereignty and natural philosophy have co-occurred under the same jurisdiction of governance. In the classic two volume work, *The Open Society and its Enemies*, the philosopher of science Karl Popper [46] defends liberal democracy against both fascism and communism by disarticulating the ideological commitments to historicism found in the latter two systems. The notion that history follows pre-determined trajectory toward some final, perfected, and inevitable political end-state is antithetical to both the notion of political liberalism and to the necessity for a free and open society. Parsons [47] suggested that a society's social structures shape its scientific development, while Barber [48] suggested that the values of democracy and science are compatible with each other and incompatible with autocratic rule. Merton [49, 50] suggested that a scientific enterprise concerned with the pursuit of impersonal truths should be unfettered by ideological dogma, is consonant with democratic governance, and clashes with totalitarianism. Work in this conceptual vein suggests a fundamental compatibility between the normative elements of democracy and science.

As Taylor [51] suggested, "[i]nnovation is inherently political in that it involves highly contested decisions over the allocation of resources, institution and policy design, and the formation and maintenance of domestic and international political economic networks" (p.150). Wang et al. [52] conducted a test of Popper's hypothesis regarding positive effects of democracy on innovation, using patents and trademarks. They show significant positive effect of democracy on innovation across a variety of empirical tests with the greatest returns occur for countries with lower levels of economic development. These results are contradicted by Gao et al. [53], who test the hypothesis on a subset of patents. Thus, the results of this literature are somewhat mixed. While these observations refer primarily to technology, it is reasonable to suggest that political governance also has significant effects on scientific innovation. In one empirical article specifically on the topic, Whetsell et al. [11] recently provided evidence in support of the democracy-science hypothesis, showing positive significant effects of democratic governance on scientific performance. They suggested that democratic governance might enhance the internal and external complexity of the country, such that science is provided wider latitude for self-organization on intellectual fitness landscapes than autocratically governed countries [54, 55]. International collaboration as a type of external complexity may produce a greater variety of conventional, novel, and atypical ideas, *c.f.* [56, 57]. Whetsell et al. [11] also showed a negative interaction effect between democracy and international collaboration intensity, where the effect of democracy on scientific performance diminished for higher levels of international collaboration intensity. This counter-intuitive findings incites further research into the effects of democracy on international research collaboration.

## Democracy and Academic Freedom

To more clearly articulate a potential causal pathway from political governance to collaborative self-organization, a newly developing empirical literature on democracy and academic freedom is briefly reviewed. Coppedge et al. [1, 2] suggest liberalism is based on a negative concept of liberty that is realized when the political power of the executive is constrained by effective checks and balances, the rule of law, constitutionally protected civil liberties, and an independent judiciary. At the same, liberalism must be combined with a structural view of electoral democracy, under which suffrage must be extensive, there must be freedom of association and expression, and there are clean elections for public officials. This structural element is based on Dahl's [58, 59] concept of *polyarchy*. However, liberal democracy is not the system of government for the majority of the world. Roughly 70% of the global population live under closed or electoral autocracies [2, 4]. As Lührmann and Lindberg [60] suggested, the world is experiencing a "third wave" of autocratization. Further, as several observers have noted, the latest wave of autocratization is less blunt and more sophisticated. In other words, autocratization is now less likely to occur as a result of an outright military coup and more likely to involve legalistic machinations, misinformation/disinformation campaigns, toxic political polarization, and sham elections [4]. As Hellmeier et al. [61] point out, autocratizing countries appear to follow a predictable pattern which typically begins by repressing academic freedom, media, and civil society. For instance, Enyedi [62] details the methods by which academic freedoms have been eroded in recent years in Hungary coinciding with its descent into electoral autocracy. According to theory, autocratization of nations should hinder the self-organization of scientists within and across national borders.

In summary, theory suggests that liberal democracy provides the optimal conditions for the self-organization of science as a whole through the development of academic freedom, where members of the academic community retain individual and collective rights to develop and communicate knowledge, are autonomous from the state and other political forces, and are free to cooperate with others across boundaries and borders [63]. Berggren and Bjørnskov [64] show that democratic political institutions are a significant predictor of academic freedom. Thus, there is a clear connection between democratic governance and the free conduct of science. In liberal democracies, where scientists are permitted and encouraged to freely associate with others across ideological, political, and national borders, we should expect greater levels of international research collaboration. In more autocratic systems, where academics are censored, monitored, and their movements are restricted, we should expect lower levels of international research collaboration. As such, the following hypothesis is warranted.

> *Hypothesis 1: Democratic governance is positively associated with international research collaboration a) tie formation and b) tie strength in the global scientific collaboration network.*

## Assortative Mixing Effects of Governance Structure

Assortative mixing refers to a process within networks in which actors tend to form connections with actors of a similar (homophily) or dissimilar (heterophily) characteristic [65]. When actors in a network form social ties based on similarities, such as race and ethnicity, religion, group membership, or any other construct real or imagined, this is often referred to as homophily in the social network literature. A common metaphor used to describe this process is "birds of a feather flock together" [66]. In contrast to homophily, when actors in a social network form connections based on complementarity or dissimilarity on a given attribute this is referred to as heterophily, and includes examples such heterosexual mate selection, mentees seeking out older "wiser" mentors, or constructing skill complementarity in team structures. In the context of the present study, the question may be formulated as follows. Are countries with similar governance structure more likely to collaborate with each other?

The question of democratic collaboration is perhaps best situated within a stream of international relations theory loosely referred to as liberal institutionalism. In contrast to realist and neorealist theories of international relations which deprioritize the internal characterization of a nation's governance structures and instead prioritize its military capabilities in balance with those of respective competitors [67, 68], liberal institutionalism stresses the importance of reducing the transaction costs of cooperation through the construction of international regimes [69]. As the logic goes, conflict can be reduced by erecting international institutions, democracies tend to engage in institution building, therefore democracies tend to cooperate more with each other and have less conflict. This vein of theory is more broadly known as the "democratic peace", the "liberal peace", or the "Kantian peace". Within the framework of liberal institutionalism, Wagner's [22] "new invisible college", as a global network of science, might be considered a type of international institution that promotes international cooperation and generally reduces the prospect of conflict between nations. These broader effects however are beyond the scope of this article. As Wagner et al. [24] suggested, the global network of science should be conceptualized as a new type of institution. It is reasonable then to question whether democracies may be more engaged in homophilous collaboration in this new institution.

The broad question posed in this article, how do the domestic characteristics of a country affect its tendencies to engage in international research collaboration, is quite like a question that political scientists have been interested in for decades. Lai and Reiter [70] argued, "the central question of international relations in the past decade has been, What are the connections between domestic and international politics"(p. 204). Leeds [71] argued that democratic leaders, by virtue of their accountability, are more likely to make and uphold credible foreign policy commitments. According to this theory, democracies might be viewed as more reliable partners for international cooperation and may also less likely to form weaker agreements with other non-democratic countries. As Gallop [15] suggests, "it seems both plausible and empirically justified that states should gain somewhat more utility when they cooperate with like regimes" (p.315). The logic of international cooperation based on regime similarity appears to be isomorphic with the research question of international research collaboration based on governance structure. However, an important difference here is that the interactions between countries discussed above tend to be formal agreements between governments, while international collaboration is the outcome of self-organization of scientific activity.

As Mansfield et al. [72, 73] show, countries with democratic governance are more likely to cooperate on trade policy. In contrast, Lai and Reiter [70] provided evidence against the hypothesis that democracies tended to engage in more formal alliances with each other. However, they did show that democracies were more likely to form defense pacts after 1945, and that, more broadly, countries with similar regime type are more likely to cooperate with one another in the international system. Using inferential network analysis techniques on a data set of international agreements between WWII and 1980, Kinne [14] shows that international cooperation is significantly driven by endogenous network effects, such as preferential attachment and transitivity. He also shows significant positive effects of democracy, which appear to overwhelm the effects of military capabilities on military agreement formation. However, interestingly he shows a negative direct effect on formal science agreements during this time period and a non-significant and positive homophily effect. In a subsequent study, Kinne [74] shows positive significant direct effects and homophily effects of democracy on diplomatic tie formation. In a similar vein, Warren [75] also uses inferential network analysis to show that democracies tend to form alliance ties with one another.

> *Hypothesis 2: Countries with similar political governance are likely to a) form and b) have stronger international research collaboration ties in the global scientific collaboration network.*

# Methods

## Modeling Approach

This article employs inferential social network analysis techniques in the family of models referred to as exponential random graph models (ERGM) [76-78]. Exponential random graph models are appropriate to account for dyadic dependencies found in relational international collaboration data. While there are numerous studies on research collaboration, fewer have employed ERGMs. However, a number of ERGM studies of collaboration have recently emerged [79-84]. In a nutshell, the ERGM provides estimates on parameters of interest by simulating a probability distribution of thousands of networks then comparing elements of the observed network to the simulated distribution. In addition to including exogenous parameters of interest, ERG models also include endogenous terms that account for complex dependencies in the data, such as density, preferential attachment, and transitivity, which are known to affect tie formation in networks. Two specific types of ERG model are used to address the hypotheses: bootstrapped temporal exponential random graph models (TERGM) are used to assess tie formation, while valued exponential random graph models (VERGM) and used to assess tie strength.

First, to test hypothesis 1a and 2a, bootstrapped TERGMs are used to account for temporal dependence in longitudinal network data [85, 86]. The TERGM uses information from prior network time points to account for subsequent network configurations, incorporating special endogenous temporal terms that account for linear time-trends and dyadic stability over time [86]. Significance tests for the parameters of interest are generated through a confidence interval approach using bootstrapped pseudolikelihood [87, 88]. The data in the current study have discrete cross-sections for the international research collaboration network between 2008 and 2017. The TERGM approach may be preferable to the stochastic actor oriented (SAOM) model in this specific case, because co-authorship ties are non-durational instances measured temporally at the year in which the paper was published. Thus, there is no actor-oriented decision to dissolve the tie in the following year, which the SAOM model assumes [89].

Second, valued ERG models (VERGMs) are used to test H1b and H2b. There currently exist limited options available to analyze edge(tie) weights in valued networks. Because the ties in the international research collaboration network are collapsed on countries as nodes, ties are stacked according to the number of instances where countries have authors collaborating with each other each year. These frequencies are referred to as edge(tie) weights or valued ties in a weighted/valued network. Edge weights are important to understand in international research collaboration because much of the variability in the data is on the weights between countries rather than the presence or absence of ties. Valued networks remain less well understood and developed and also pose greater computational challenges than binary ERGMs [90]. Often, such networks are simply binarized and analyzed using standard ERG models [87]. One method to binarize valued networks based on tie weights is a disparity approach to estimating statistical significance of ties using the R package 'backone' [91, 92]. More recently, VERGMs have been advanced and implemented in the R package 'ergm.count' to analyze valued networks [93, 94], which is part of the broader R package 'statnet'[95]. Rather, than focusing on the probability of a tie forming between two nodes, VERGMs produce estimates based on the strength of ties. Thus, VERGMs are suitable to addressing the hypotheses regarding the strength of research collaboration ties in the international system. As Huang and Butts [90] suggested, there is an "acute" need for analysis of the strength of social connections between nations. In order to test for effects over time, VERGMs are replicated for each yearly network cross section from 2008 to 2017. More details on the implementation of the models are described below in the results section.

## Data and Variables

This research combines data from numerous sources. First, international research collaboration metadata were gathered from Indiana University's CADRE portal containing the Web of Science database on all articles and reviews for the years 2008 to 2017 [96]. The metadata includes author-country affiliations for individual papers. These data are aggregated at the country level, where papers represent network ties between two or more affiliated countries. Under this construction, the network is a symmetrical adjacency matrix, M, where M$_{ij}$ equals the number of times a paper published each year includes an author from country $i$ and an author from country $j$. Aggregating at the country level produces a valued network where all ties between the countries are aggregated as frequency weights on the network ties. There are at least three reasons for aggregating at the country-level, as opposed to using author-author network data. First, ERGMs tend to be computationally infeasible with extremely large networks, i.e. the global co-author network for all of science comprises hundreds of thousands if not millions of nodes. Although, work on this problem appear to be advancing [97]. Second, disambiguation of names remains a persistent problem in studies of co-author networks, where names are entered differently across publications or multiple individuals have identical names. Treating countries as nodes avoids the name disambiguation problem because a network tie(s) is created between two or more countries when a paper includes more than one country affiliation, independent of how the author lists their name or whether their name is ambiguous, *c.f.* [98]. Third, the primary independent variable (liberal democracy) is measured at the country level, such a democracy score applied to individuals would be invariant within each country-year unless individuals change affiliations. While such an analysis would be an important contribution, it is not within the scope of the present article. Since there are already a large proportion of binary research collaboration ties between many countries in the international system, much of the variability in the data resides in the edge weights. The data were used to produce symmetrical, undirected, valued matrices for each year between 2008 and 2017. The exact number and order of nodes across years was maintained for comparability between years using the final year (2017) as the node set, resulting in a sample of 200 countries. However, after eliminating country column/rows for those with missing data in the primary independent variable, the final node sample included 170 countries.

The primary independent variable of interest is the Varieties of Democracy Institute's (V-Dem) liberal democracy index [1, 2]. The V-Dem data represents the most comprehensive and current characterization of democracy across nations in the international system and is more granular than other extant measures such as the Economist Democracy Index or the Polity project. The liberal democracy index combines an index of liberalism with an electoral democracy index. The liberal component is composed of three sub-components: 1) equality before the law and individual liberties, 2) judicial constraints on the executive, and 3) legislative constraints on the executive. The electoral democracy index is based on Robert Dahl's [58, 59] characterization of polyarchy, and is composed of five democracy components measuring 1) freedom of association, 2) free and fair elections, 3) freedom of expression and alternative sources of information, 4) elected official, 5) and suffrage [2]. The index is standardized on a 0-1 scale, where higher on the scale represents a greater realization of the "ideal" of liberal democracy.

Fig 1. Liberal Democracy among 170 Countries 2008-2017

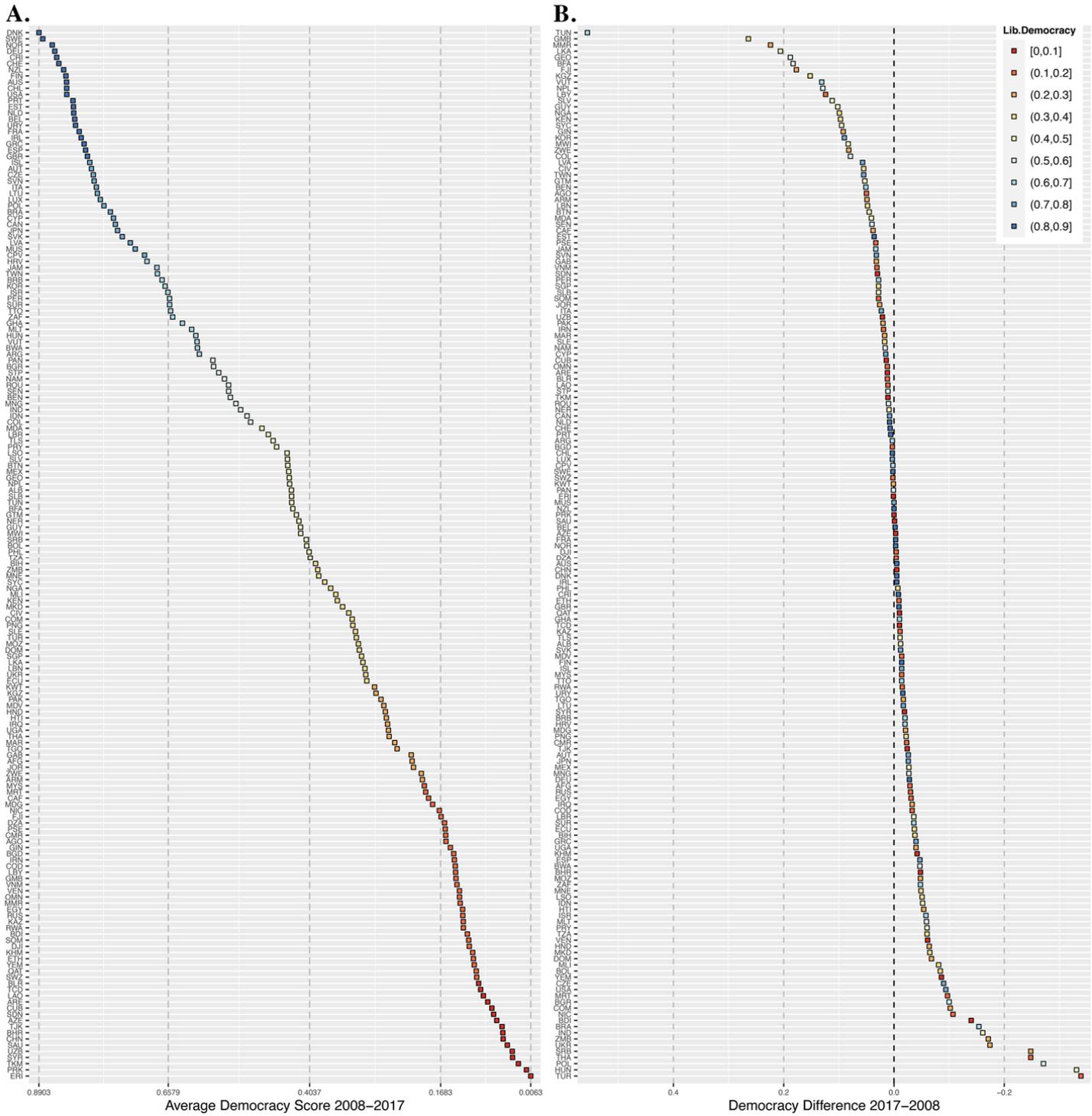

Fig 1 shows two plots: Fig 1a and the left shows the time frame average liberal democracy score of each country listed in order from highest to lowest score, Fig 1b shows the difference in liberal democracy score between 2017 and 2008 list in order of largest increase in score to largest decrease in score. In Fig 1b countries are colored by their average liberal democracy score, while in Fig 1b countries are colored by their 2017 score. Notably no country scored higher than 0.9 on the index. Fig 1b shows many less developed countries such as TUN, GMB, MMR, LKA, GEO, BFA, FJI, KGZ, VUT, NPL, and LBY had the greatest increases in democracy, while more developed countries like TUR, HUN, POL, UKR had the greatest declines. Meanwhile, developed countries like FRA, GBR, and CHN have virtually no change, while USA declined slightly. However, end point differences of course mask non-linear changes that occurred between 2008 and 2017.

This research employed a select number of control variables to account for systematic differences between countries that might also drive international research collaboration. These include three measures from World Bank Indicators, including the natural logarithm of gross domestic product per capita, the natural logarithm of population size, and level of urbanization. The inclusion of these variables as controls is relatively straightforward. The wealth and population size of the nation likely enhances its ability to participate in a variety of collaborative scientific activities, while urbanization may reduce the need to collaborate externally due to a concentration of domestic collaboration partners[99]. The natural logarithm of the number of authors, aggregated at the country level, was used as a measure of national scientific capacity to control for higher tie probability due to larger numbers of scholars available for collaboration. This variable may provide a more granular control than other commonly used measures such as number of Ph.D. granting institutions or secondary school enrollment. These data were captured separately from Elsevier's Scopus data. Other control variables were considered such as R&D/GDP and gross domestic expenditure on research and development (GERD). However, the data coverage for these variables is far smaller and therefore were not included due to the broad nature of the hypotheses. Finally, it is important to control for geography since policy and proximity effects are known to affect collaboration[80, 100, 101]. Geography was controlled for in two ways: 1) region was included as a fixed factor to account for proximity and political effects of geography by using the V-Dem data's geo-political region variable, which includes ten regions; and 2) a country-country distance matrix was used as an edge covariate. The distance matrix was generated from the R 'cshapes' package. Distances were calculated by closest border-to-border distance using the Correlates of War (COW) coding of countries. Lastly, homophily/heterophily terms were also included for all node attribute controls.

    Network models in the exponential family offer a wide variety of endogenous terms which model the various sources of dependency in relational data. Scholarship using ERGMs has produced a relatively standard set of model terms that tend to be used to account for social processes, such as density (edges), preferential attachment (gwdegree – geometrically weighted degree distribution), and transitivity (gwesp – geometrically weighted edgewise shared partner distribution). For VERGMs, a different logic is applied. VERGM uses the term "sum", which is similar to the term "edges". The sum term represents the intercept and is used to account for the endogenous effect of tie strength. The term "nodesqrtcovar" is used to account for actor heterogeneity in tie strength (a proxy for preferential attachment). The term "transitive.weights" is used to account for transitivity with the min, max, min specification. The transitive.weights term is considered a hierarchical transitivity term. Following, Pilny and Atouba [102], nodesqrtcovar and transitive.weights are interpreted similarly to gwdegree and gwesp respectively. Lastly, the model term "nonzero" that estimates the amount of zero inflation is included in the VERGMs.

Some variables had missing data. ERG models require no missing data on node covariates. Thus, the node file and the network matrixes must contain complete data on the same set of nodes. First, the V-Dem data was used as the basis for the data merge within the node file, which contained data on liberal democracy for 179 countries. Second, after merging all other data with the V-Dem data, the countries SSD, SML, XKX, and ZZB were removed for completely missing data on the controls. Third, the V-dem data also contained two country codes for Palestine: PSE appears to be the most used, so PSG was dropped. This left 6 countries with missing data: ERI, PRK, SOM, SYR, TWN, and VEN. Therefore, the following imputation strategy was taken. Fourth, TWN, VEN, and PRK did not contain World Bank GDP data, which were available in the V-Dem data and were imputed manually. TWN did not contain data on population size or urbanization from the World Bank and were instead gathered from https://www.statista.com/ and imputed manually; urbanization data was partially missing across 5-year intervals and was carried forward across missing years. Fifth, ERI, SOM, and VEN had partially missing data on GDP, population size, and urbanization, and so their values were imputed from the average of the existing country data. Sixth, at this point the node file was matched to the matrixes, which dropped COG, GNB, GNQ, HKG from the V-dem Data, and 30 countries were dropped from the matrixes (AND, ANT, ATG, BHS, BLZ, BMU, BRN, FSM, GIB, GLP, GRD, GRL, GUF, KNA, KOS, LCA, LIE, MCO, NCL, PLW, PYF, REU, SMR, SSD, TOA, VAT, VCT, WAS, WSM, ZRN). This left 170 countries matching in both the node file and the matrixes. Finally, the country-country distances for Israel were inputted in the distance matrix for Palestine because the geographical distance data did not include Palestine as a country.

# Results

The results are presented as follows. First, visualizations of the networks are presented to illustrate the evolution of the network over the time frame. Second, descriptive statistics on the structure of the whole network are presented for each year. Third, temporal exponential random graph models (TERGM) are presented that show tests of hypotheses regarding the formation of international research collaboration ties over time. Finally, valued exponential random graph models (VERGM) are presented showing the hypotheses tests regarding the strength of international research collaboration ties.

**Fig 2. Trimmed International Collaboration Network 2008 to 2017**

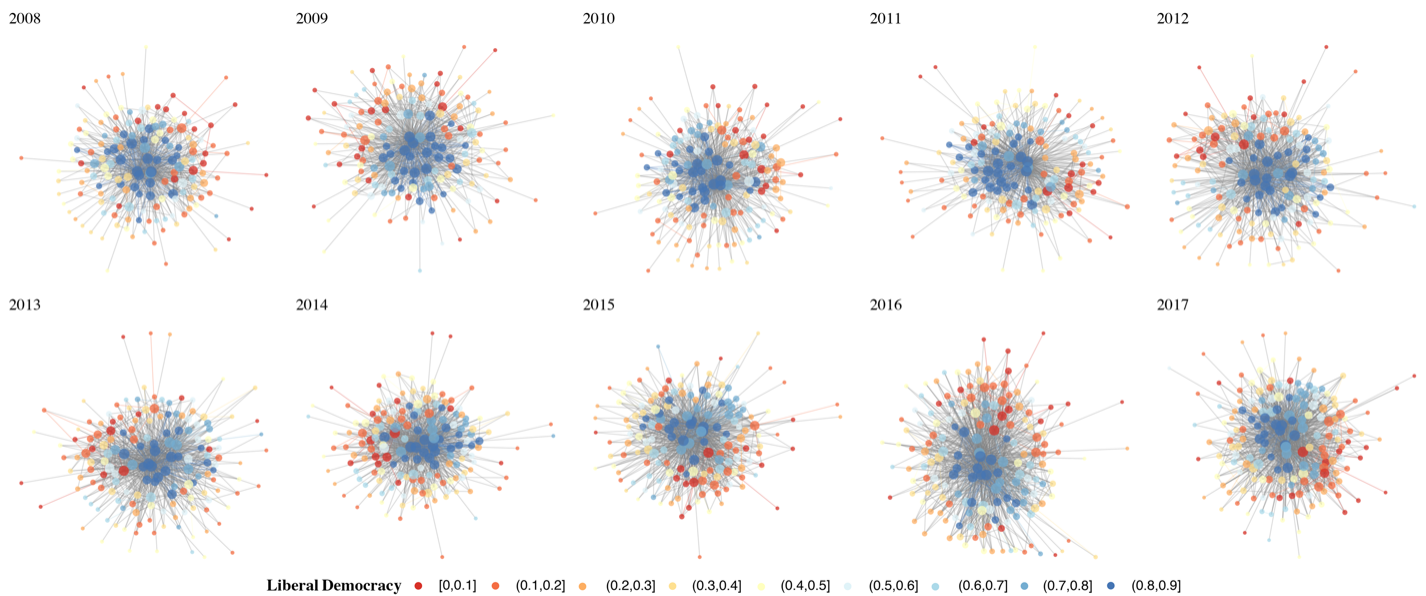

Fig 2 shows the international collaboration networks left to right by year on the top row from 2008-2012 and 2013-2017 on the bottom row. The R package 'backbone' was used to reduce the density (Trimmed) of the networks for visualization purposes [91]. This package implements a disparity filter that compares observed edge(tie) weights to those in a simulated null model retaining edges that are statistically significant at a specified level. At the 5% significance level, the procedure reduced the number of edges in the networks between 85 to 87 percent across networks-years. For both Fig 2 and Fig 3, the nodes are colored using the R 'RdYlBu' palette, with the level of liberal democracy corresponding to the color indicated in the figure legend. Isolates are removed from the figure for visualization purposes. The Fruchterman-Reingold algorithm [103] was used as the network layout, nodes are sized by degree centrality, and the R package 'GGally' was used to visualize the network [104]. For comparison purposes Fig 3 shows full untrimmed networks with edge weights, corresponding to the whole network descriptive statistics presented below. Both trimmed and untrimmed networks show that countries scoring lower on the liberal democracy index tend to occupy peripheral positions in the network, while those scoring higher tend to occupy the core of the network.

**Fig 3. Untrimmed International Collaboration Network 2008 to 2017**

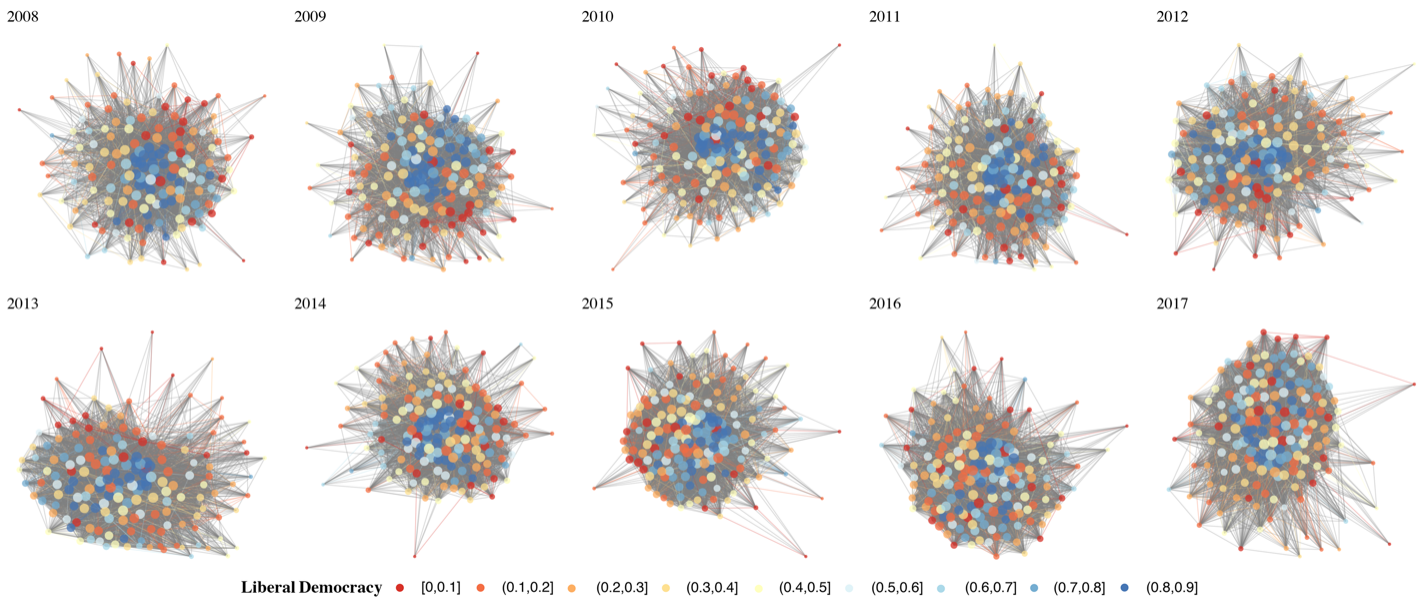

Table 1 shows the whole network descriptive statistics for each year of the network from 2008 to 2017. The total number of nodes (countries) is the same for each cross section. The total number of edges, total edge weight, and maximum edge weight increase continuously over time. The number of disconnected components decreases over time. One very striking finding is the substantial increase in the density of the network over the time frame.

**Table 1. Whole Network Descriptive Statistics 2008 to 2017**

|                | 2008   | 2009   | 2010   | 2011   | 2012   | 2013   | 2014    | 2015    | 2016    | 2017    |
|----------------|--------|--------|--------|--------|--------|--------|---------|---------|---------|---------|
| nodes          | 170    | 170    | 170    | 170    | 170    | 170    | 170     | 170     | 170     | 170     |
| isolates       | 13     | 9      | 8      | 6      | 5      | 4      | 3       | 3       | 3       | 0       |
| total edges    | 5475   | 5843   | 6173   | 6417   | 7174   | 7330   | 8153    | 8844    | 9659    | 10100   |
| total weight   | 507288 | 525745 | 631924 | 764410 | 914119 | 966903 | 1071010 | 1195662 | 1400176 | 1427240 |
| max weight     | 13593  | 14543  | 15577  | 18771  | 22059  | 26676  | 30882   | 35906   | 40900   | 43706   |
| components     | 14     | 10     | 9      | 7      | 6      | 5      | 4       | 4       | 4       | 1       |
| density        | 0.381  | 0.407  | 0.430  | 0.447  | 0.499  | 0.510  | 0.568   | 0.616   | 0.672   | 0.703   |
| centralization | 0.533  | 0.537  | 0.514  | 0.521  | 0.468  | 0.463  | 0.411   | 0.357   | 0.312   | 0.299   |
| closed triads  | 108845 | 121879 | 135294 | 141073 | 183467 | 187762 | 239457  | 283219  | 341552  | 372398  |

Table 1 Notes: The table shows descriptive statistics of network cross-sections. Nodes is the number of countries. Isolates is the number that have no connections. Total edges is the sum of ties between countries. Total weight is the sum of edge weights between all countries. Components is the number of clusters disconnected from the main cluster (1 means all clusters are connected). Density is the proportion of observed to potential ties (unweighted). Centralization is the degree to which ties are centered on few nodes (unweighted). Closed triads is the number of closed triads in the network (unweighted).

The network becomes extremely dense by 2017, which presents computational challenges for inferential network models. At the same time the centralization of the network is declining over time as more countries are connecting to each other and not just to a few highly connected countries. Finally, among the 804,440 possible closed triads in the network (sum of undirected triad types listed in 'triad.census' function), the triad census shows an incredible proportion (50%) of closed triads by 2017, more than tripling from 108,854 to 372,398 over the time frame. This may be due, at least in part, to the nature of multi-country publications, which by definition are represented as a set of closed triads between all countries listed on a publication.

Before discussing the results of the bootstrapped TERGMs, there are a few important notes on the modeling approach. Because TERGMs are unable to account for edge weights in valued networks, four separate models are presented that binarize (0/1) the edge weights at certain cut points. In the first model, edge weights of zero are left zero, and edge weights at one or higher are set to one. This is termed the "full network", as this does not affect the number of ties or the density in the network. This is a common strategy for addressing edge weights and density issues in network analysis [87]. Next, in models two, three, and four, the networks are binarized using the disparity procedure in the R 'backbone' package which trims edges based on the statistical significance of the edge [91]. Model 2 uses a 50% significance level, Model 3 uses a 25% percent significance level, and Model 4 uses a 5% significance level to trim the edges. Binarizing in this way effectively reduces the density of the networks because the ties require a progressively higher edge weight values to be counted as ties in the network. Table 2 shows the TERGM results of the four models.

**Table 2. Temporal Exponential Random Graph Models 2008-2017**

|  | **Model 1: All Ties** | **Model 2: 50% Edge Sig.** | **Model 3: 25% Edge Sig.** | **Model 4: 5% Edge Sig.** |
|---|---|---|---|---|
| Liberal Democracy | 0.616 [0.422, 0.835] * | 0.229 [0.072, 0.384]* | 0.120 [0.014, 0.216]* | 0.118 [-0.091, 0.337] |
| Dem. Heterophily | -0.462 [-0.630, -0.301]* | -0.252 [-0.389, -0.142]* | -0.107 [-0.230, 0.026] | -0.398 [-0.647, -0.172]* |
| Num. Authors | 0.557 [0.521, 0.599]* | 0.419 [0.382, 0.456]* | 0.491 [0.445, 0.540]* | 0.508 [0.436, 0.583]* |
| Auth. Heterophily | -0.163 [-0.205, -0.126]* | 0.128 [0.119, 0.138]* | 0.196 [0.156, 0.236]* | 0.388 [0.349, 0.424]* |
| Urbanization | -0.004 [-0.006, -0.001]* | 0.000 [-0.002, 0.002] | -0.003 [-0.005, -0.001]* | -0.003 [-0.007, 0.001] |
| Urb. Heterophily | -0.002 [-0.004, 0.000] | -0.003 [-0.006, 0.001] | -0.004 [-0.006, -0.001]* | -0.006 [-0.011, -0.002]* |
| GDP Per Capita | 0.124 [0.101, 0.148]* | 0.067 [0.024, 0.113]* | 0.110 [0.037, 0.193]* | 0.226 [0.161, 0.291]* |
| GDP Heterophily | -0.094 [-0.124, -0.065]* | -0.100 [-0.139, -0.054]* | -0.037 [-0.085, 0.012] | 0.125 [0.042, 0.202]* |
| Population Size | 0.132 [0.070, 0.197]* | 0.117 [0.029, 0.201]* | 0.072 [-0.014, 0.146] | 0.127 [0.046, 0.207]* |
| Pop. Heterophily | -0.006 [-0.035, 0.026] | -0.073 [-0.111, -0.034]* | -0.095 [-0.132, -0.059]* | -0.149 [-0.195, -0.100]* |
| Region | FIXED | FIXED | FIXED | FIXED |
| Reg. Homophily | 1.959 [1.777, 2.156]* | 2.166 [1.985, 2.371]* | 2.105 [1.926, 2.283]* | 1.865 [1.729, 2.008]* |
| Geo. Distance | -1.548 [-1.759, -1.363]* | -1.533 [-1.681, -1.359]* | -1.762 [-1.904, -1.645]* | -1.874 [-2.074, -1.668]* |
| Density | -437.460 [-916.941, -184.433]* | -12.240 [-16.405, -8.282]* | -12.853 [-15.770, -9.479]* | -19.071 [-21.644, -16.690]* |
| Pref. Attachment | -559.040 [-1146.193, -241.540]* | -5.932 [-7.484, -4.849]* | -2.005 [-2.657, -1.279]* | -0.527 [-0.881, -0.168]* |
| Transitivity | 329.750 [132.906, 702.778]* | -0.090 [-1.108, 1.362] | 0.087 [-0.386, 0.614] | 0.150 [-0.121, 0.461] |
| Time Autoreg. Lag | 0.747 [0.716, 0.781]* | 1.517 [1.442, 1.597]* | 1.889 [1.834, 1.954]* | 2.434 [2.395, 2.472]* |
| Time Dependence | 0.169 [0.122, 0.224]* | 0.110 [0.066, 0.167]* | 0.069 [0.033, 0.102]* | 0.038 [-0.005, 0.105] |

Table 2 Notes. Estimates are bootstrapped mean. Confidence intervals in brackets. *= "significant" when confidence intervals do not contain 0. Model 1 includes all edges, which are binarized to 1 if edge weights >= 1. Model 2 binarizes edges if weights are significant at the 50% level using the disparity procedure. Model 3 is the 25% level. Model 4 is the 5% level. Negative significant estimates on heterophily terms indicate

homophily. Positive on homophily terms indicate homophily. 'Density' is the 'edges' term. 'Pref. Attachment' is 'gwdegree' with a fixed alpha of 0.5 for all models (negative indicates preferential attachment). 'Transitivity' is 'gwesp' with a fixed alpha of 0.25 for all models. 'Time. Autoreg. Lag' is an autoregressive edge covariate term with a 1 year lag. 'Time Dependence' represents a linear time trend. All models were run on 1500 simulations per year, using parallel 'snow' on 5 cores.

The variable Liberal Democracy provides support for hypothesis 1(a) across models 1-3, showing that the liberal democracy index has a positive significant estimate (bootstrapped mean) on the probability of tie formation in the network. This term is not "significant" in Model 4, but the estimate remains positive and upper confidence interval is almost four times higher than the lower confidence interval is low. Note that estimates are "significant" if the confidence intervals in brackets do not contain zero. The results also show support for H2a, regarding governance homophily effects on tie formation. The model term Dem. Heterophily shows significant negative coefficients across models 1,2, and 4, suggesting ties tend to form between countries with similar liberal democracy scores. Model 3 is not significant, but the estimate remains negative, and the lower confidence interval is almost nine times lower than the upper is high. The negative sign indicates that a positive absolute difference is associated with a decrease in the formation of ties (heterophily), while a negative absolute difference is associated with an increase in collaborative ties (homophily).

Changes across the four models suggest that as the network is trimmed with an elevated criterion for the counting of ties (higher edge weight significance necessary for counting the tie) the size of the estimate decreases. In other words, a portion of the magnitude of the effect of Liberal Democracy on tie formation appears to be accounted for by tie formation between less well developed and less scientifically prominent countries. The larger the sample of network ties included in the model, the greater the strength of the estimate on Liberal Democracy. This could also be due in part to a lower tie change rate for higher edge-weight thresholds.

For exogenous controls, the estimates on Num. Authors is positive and significant across all models. Auth. Heterophily shows a different pattern with significant negative effects in the full network (Model 1) but reversed effects in the trimmed networks (Model 2-4), suggesting that more scientifically established nations tend to collaborate with countries of differing author capacity. Urbanization is mostly negative across models, while Urb. Heterophily is negative but reverses and becomes significant in Model 4, indicating homophily based on level of urbanization is a driver of ties in the restricted network. GDP Per Capita has positive significant effects, and GDP Heterophily is negative and significant in 1 and 2 but reverses in 4, indicating collaboration at higher strength thresholds tends to occur between countries of differing GDP. Geography shows a strong consistent effect on tie formation. Region was included as a fixed effect, while Reg. Homophily showed positive significant estimates as collaborative ties tend to form within regions. The edge covariate Geo. Distance shows a significant negative effect, which indicates that tie formation tends to decrease the further away countries are from each other.

The endogenous controls also show interesting results. Pref. Attachment shows significant negative coefficients across the models (indicating preferential attachment), while transitivity appears to lose significance at stricter trimming levels. Lastly, Time Autoreg. Lag is positive and significant in all models suggesting ties tend to carry over to subsequent years, and Time Dependence indicates the presence of a linear time trend in Models 1-3 but not 4.

The bootstrapped TERGMs converged and produced reasonable MCMC diagnostic results and goodness-of-fit statistics for edge-wise shared partner distribution, degree distribution and geodesic distance, illustrated in Fig 4 corresponding to Models 1 through 4 in Table 2. A reasonably fitting model will tend to show the observed statistic from the network analysis represented by the black line contained within the error bars representing the simulated distributions. All four models show a good fit. Multi-collinearity diagnostics suggested low variance inflation among all model terms except some variance inflation for Model 1 on the endogenous network controls [105].

Fig 4. Goodness-of-fit Statistics Plots for TERGMs.

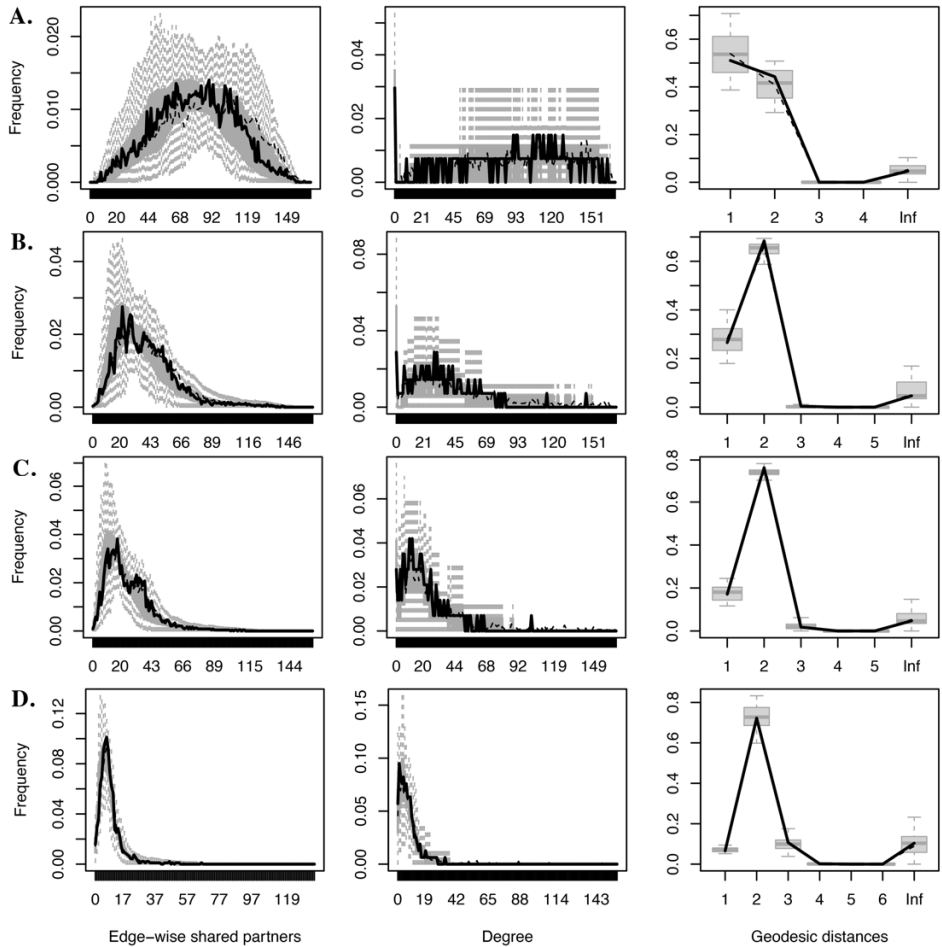

Next, the results of the valued ERGMs (VERGM) are presented in Table 3. To achieve convergence in the VERGMs, the edge weights (+1) were first transformed using the natural logarithm then quantized (assessed without zeros), resulting in a range of edge values between 0 and 5. The binomial reference distribution was selected for the models. Three valued network controls were included: Sum Weight is the intercept term. Zero Inflation accounts for presence of zero values in the network. Pref.Attachment estimates the propensity of countries to form stronger ties based on their existing tie strength. As a side note, a transitivity term that estimates the propensity of country sets to form closed triads, was extensively tested in the models. However, the term appeared to induce severe multicollinearity in the endogenous and exogenous model terms, its inclusion did not significantly alter democracy effects, it lacked statistical significance in several of the models, and it slowed computation time exponentially.

**Table 3: Valued Exponential Random Graph Models 2008-2017**

|  | 2008 | 2009 | 2010 | 2011 | 2012 | 2013 | 2014 | 2015 | 2016 | 2017 |
|---|---|---|---|---|---|---|---|---|---|---|
| **Liberal Democracy** | 0.302*** | 0.300*** | 0.241*** | 0.178*** | 0.108* | 0.078 | 0.104* | 0.079 | 0.265*** | 0.244*** |
|  | (0.050) | (0.050) | (0.050) | (0.047) | (0.045) | (0.044) | (0.043) | (0.043) | (0.039) | (0.033) |
| **Dem. Heterophily** | -0.385*** | -0.484*** | -0.386*** | -0.335*** | -0.291*** | -0.223*** | -0.199*** | -0.138* | -0.225*** | -0.180*** |
|  | (0.057) | (0.059) | (0.059) | (0.058) | (0.057) | (0.059) | (0.055) | (0.057) | (0.052) | (0.048) |
| **Num. Authors** | 0.340*** | 0.347*** | 0.405*** | 0.404*** | 0.432*** | 0.438*** | 0.443*** | 0.432*** | 0.321*** | 0.224*** |
|  | (0.015) | (0.015) | (0.015) | (0.016) | (0.015) | (0.015) | (0.014) | (0.013) | (0.012) | (0.011) |
| **Auth. Heterophily** | -0.069*** | -0.090*** | -0.104*** | -0.104*** | -0.107*** | -0.103*** | -0.128*** | -0.149*** | -0.163*** | -0.090*** |
|  | (0.008) | (0.008) | (0.007) | (0.007) | (0.007) | (0.006) | (0.006) | (0.006) | (0.006) | (0.005) |
| **Urbanization** | -0.000 | -0.001 | -0.001 | -0.003*** | -0.002* | -0.003*** | -0.001 | 0.000 | -0.002** | -0.002*** |
|  | (0.001) | (0.001) | (0.001) | (0.001) | (0.001) | (0.001) | (0.001) | (0.001) | (0.001) | (0.000) |
| **Urban Heterophily** | 0.000 | 0.000 | -0.001 | 0.001 | -0.001 | 0.001 | 0.002** | 0.001 | -0.000 | 0.000 |
|  | (0.001) | (0.001) | (0.001) | (0.001) | (0.001) | (0.001) | (0.001) | (0.001) | (0.001) | (0.001) |
| **GDP Per Capita** | 0.077*** | 0.115*** | 0.083*** | 0.095*** | 0.071*** | 0.106*** | 0.097*** | 0.095*** | 0.114*** | 0.047*** |
|  | (0.017) | (0.018) | (0.017) | (0.017) | (0.015) | (0.015) | (0.015) | (0.014) | (0.013) | (0.010) |
| **GDP Heterophily** | -0.024 | -0.047** | -0.049** | -0.079*** | -0.081*** | -0.111*** | -0.121*** | -0.135*** | -0.055*** | -0.050*** |
|  | (0.015) | (0.015) | (0.015) | (0.015) | (0.015) | (0.015) | (0.015) | (0.014) | (0.012) | (0.010) |
| **Population Size** | 0.107*** | 0.121*** | 0.097*** | 0.077*** | 0.047*** | 0.038** | 0.044*** | 0.066*** | 0.113*** | 0.037*** |
|  | (0.013) | (0.014) | (0.013) | (0.013) | (0.012) | (0.013) | (0.012) | (0.012) | (0.011) | (0.009) |
| **Pop. Heterophily** | -0.047*** | -0.044*** | -0.052*** | -0.042*** | -0.058*** | -0.061*** | -0.051*** | -0.061*** | 0.002 | 0.005 |
|  | (0.009) | (0.009) | (0.009) | (0.009) | (0.009) | (0.008) | (0.009) | (0.008) | (0.007) | (0.006) |
| **Region** | FIXED | FIXED | FIXED | FIXED | FIXED | FIXED | FIXED | FIXED | FIXED | FIXED |
| **Reg. Homophily** | 1.416*** | 1.567*** | 1.529*** | 1.609*** | 1.494*** | 1.619*** | 1.672*** | 1.776*** | 1.485*** | 1.448*** |
|  | (0.042) | (0.042) | (0.043) | (0.043) | (0.041) | (0.042) | (0.042) | (0.041) | (0.039) | (0.037) |
| **Geo. Distance** | -1.000*** | -0.915*** | -1.101*** | -1.024*** | -1.080*** | -1.054*** | -1.152*** | -1.065*** | -1.060*** | -1.027*** |
|  | (0.069) | (0.070) | (0.074) | (0.072) | (0.077) | (0.074) | (0.079) | (0.077) | (0.082) | (0.084) |
| **Sum Weight** | -10.116*** | -11.391*** | -10.646*** | -10.139*** | -9.151*** | -9.596*** | -9.863*** | -10.540*** | -9.750*** | -4.716*** |
|  | (0.564) | (0.577) | (0.540) | (0.528) | (0.466) | (0.492) | (0.491) | (0.471) | (0.457) | (0.378) |
| **Zero Inflation** | -2.861*** | -2.540*** | -2.071*** | -2.106*** | -1.750*** | -1.808*** | -1.268*** | -0.696*** | -2.123*** | -1.941*** |
|  | (0.066) | (0.061) | (0.059) | (0.058) | (0.052) | (0.052) | (0.049) | (0.047) | (0.057) | (0.058) |
| **Pref. Attachment** | 1.115*** | 0.983*** | 1.166*** | 1.232*** | 1.407*** | 1.368*** | 1.636*** | 2.017*** | 2.021*** | 3.172*** |
|  | (0.139) | (0.135) | (0.128) | (0.131) | (0.119) | (0.119) | (0.112) | (0.109) | (0.105) | (0.106) |

Table 3 Notes. Standard errors in parentheses. ***p < 0.001; **p < 0.01; *p < 0.05. Negative significant estimates on heterophily terms indicate homophily. Positive on homophily terms indicate homophily. Sum Weight, is the 'sum' term, representing the intercept. Zero inflation is 'nonzero' representing a prevalence of zeros. Pref. Attachment is 'nodecovarsqrt'. All models were run on 15000 simulations. All models used parallel type PSOCK on 7 CPU cores.

The models presented in Table 3 provide support from 2008 to 2017 for both H1b and H2b, which specify respectively that the liberal democracy index is significantly associated with the strength of international research collaboration ties in eight out of ten network years and that similarity in the index is associated with increased strength of international collaboration ties between countries in all years. Support for the hypothesized main effect of Liberal Democracy appears to wane in the middle years then increase in strength toward the end. During years 2013 and 2015 the direct effect is not significant at the 5% significance level. However, they are significant at the 10% level with a 0.079 p-value and a 0.065 p-value respectively. Support for the hypothesized homophily effect of democracy on collaboration strength is also supported by the negative significant coefficient on Dem. Heterophily across the time frame (negative is homophily; positive is heterophily). There are also positive estimates on Num. Authors, GDP Per Capita, and Population Size. Auth. Heterophily, GDP Heterophily, and Pop. Heterophily also suggest significant homophily across the time frame. Region was included as a fixed effect. Reg. Homophily shows positive significant estimates across the models, suggesting stronger ties tend to form within regions. Geo. Distance, measured through an edge covariate distance matrix, shows a strong negative estimate, indicating that increasing geographical distance between countries reduces the propensity to form strong collaborative ties. Urbanization and Urb. Heterophily were mostly non-significant. For endogenous terms, Preferential attachment showed significant estimates that appeared to increase in strength across the time frame. There is currently no implementation of the goodness-of-fit procedure for VERGMS. Finally, the variance inflation factor (VIF) test for ERGMs [105] showed no issues with multicollinearity in the VERGMs.

***Limitations.*** There are a few limitations to the current study. While the V-Dem data does have a very broad coverage of countries in the international system, there were at least thirty countries contained in the bibliometric data which were not covered. In this case, having very broad data coverage is important because much of the variability over time is between less developed countries, and democracy effects may be underestimated. Another limitation is that there currently are no valued network options for temporal inferential network models. Another limitation was the failure to find a model to converge on the valued network with raw edge weights using the Poisson distribution. This may be due to the extreme density of the networks and the wide dispersion in the edge weight values [90]. There are also well-known limitations of the Web of Science data base used for the networks. Finally, as with all observational studies, in the absence of a control group and randomization, causal inference cannot be conclusively established regarding hypothesized effects.

## Discussion and Conclusion

This study sought out to test two hypotheses, each with two sub-hypotheses, using longitudinal data merged from the Varieties of Democracy Institute, Scopus, Web of Science, and World Bank Indicators, and analyzed with two types of inferential network analysis: bootstrapped temporal exponential random graph models (TERGM), which leveraged the panel structure of the data; and, valued exponential random graph models (VERGM), leveraging edge weights but treating each year sequentially over time.

      The first hypothesis concerned the direct effects of liberal democracy on international collaboration with significant effects across tie formation (H1a), and strength (H1b). The study results supported the hypotheses. Yet in the TERGMs these results appear to depend on the tie

inclusion criteria, with more restrictive criteria based on stronger tie weights attenuating these effects. However, the VERGM results show that liberal democracy appears to be a significant predictor of tie strength. More generally the results provide empirical support for the argument that democracy and science share similar values, are productively compatible, and that democracy provides the context for the full development and flourishing of global scientific inquiry [46-50]. The results also add to an emerging literature that broadly seeks to test the relationship between democracy and science empirically [11]. Further, the results speak to a growing body of research on autocratization and the erosion of academic freedom around the world [4, 60, 64]. Future research should investigate what impacts autocratization is having on global science.

    The second hypothesis concerned assortative mixing effects of liberal democracy on international collaboration ties with uniform effects across tie formation (H2a), and strength (H2b). The TERGM results showed significant positive results regarding the hypothesized effect of governance homophily in the formation of ties, suggesting that researchers from countries with similar levels of liberal democracy and conversely, autocracy, are more likely to form and maintain international collaboration ties. The VERGM results suggested that strong ties are more likely to form between countries of similar political governance. These results speak to a stream of international relations literature that has debated whether democracies really do cooperate with each other more or less than countries with other types of governance [14, 15]. While this literature has come to include attention to endogenous network effects in recent years, the current study shows that democracies indeed appear to be cooperating with one another on international science, despite exogenous economic, geographical, and population factors, as well as endogenous network processes such as preferential attachment and transitivity, which have often

rendered political governance non-significant. The current study shows that homophily of governance structure is an important empirical predictor of international research collaboration.

This study also uncovered several non-hypothesized but nevertheless interesting results. Economic capacity, researcher capacity, and geo-political region appear to be significant drivers of the strength of international collaboration. In addition, these factors all appear to show strong positive homophily estimates on tie formation. In other words, international collaboration ties appear more likely to form and strengthen between economically powerful countries, between countries with similar researcher capacity, and between countries in the same geo-political region. Further, geographic distance appears to be a significant impediment to tie formation and strength.

One methodological finding that may be of interest to future researchers is that studying international collaboration requires an understanding of existing network density. Currently, the network is extremely dense with almost all ties between countries being realized with at least one country-country paper tie. This presents challenges for researchers seeking to understand country level effects on international research since there is less tie change variability between well-connected countries. However, this also points to the need for greater attention to the effects of governance factors on science among traditionally less well-connected countries. Future research should further investigate the connections between factors of governance and processes and outcomes of international science. As the international collaboration network reaches near complete density, it will be necessary to investigate various sub-networks of disciplines selected based on hypothesized effects of governance factors. Further, newer measures of academic freedom have recently been developed which invites more fine-grained investigation into the details of democratic effects at the institutional level. Future studies should link up network measures used to characterize country position with scientific outcomes such as novelty and impact.

This study addresses a gap in the literature on international collaboration which has thus far almost completely avoided the question of comparative political governance. This is an unfortunate lacuna since insights about the structural antecedents of international research collaboration are relevant to policymakers seeking greater scientific impact on the world stage. While the world undergoes a third wave of democratic backsliding, the question remains open as to the long-term effects on international collaboration. This study shows that in fact liberal democracy is a significant predictor of international research collaboration. At the same time, a large majority of the world's citizens live under varying degrees of autocratic rule, where academic freedom is undermined or non-existent. Policymakers seeking to develop advanced scientific systems should consider democratic reforms that promote greater development of scientific human capital, remove constraints from academic freedom, and provide support for the self-organization of scientific activity.

# Acknowledgements

Thank you to Caroline Wagner, Xiaojing Cai, and Jeroen Bass for access to data. Thank you to Michael Siciliano, Mark Zak Taylor, Gordon Kingsley, and Brian An for comments on drafts. Thank you to Pavel Krivitsky for correspondence on VERGM model specification. Thank you to Scott Duxbury for correspondence on multicollinearity in ERGMs. Thank you to Philip Leifeld for correspondence on BTERGMs. This article has an arXiv preprint #2203.01827.

# Bibliography

1. Coppedge, M., et al., *Conceptualizing and measuring democracy: A new approach.* Perspectives on Politics, 2011. **9**(2): p. 247-267.
2. Coppedge, M., et al., *"V-Dem Codebook v12" Varieties of Democracy (V-Dem) Project.* 2022.
3. Fortunato, S., et al., *Science of science.* Science, 2018. **359**(6379): p. eaao0185.
4. Boese, V.A., et al., *Autocratization Changing Nature? Democracy Report 2022.* Varieties of Democracy Institute (V-Dem), 2022.
5. Zhou, P. and L. Leydesdorff, *The emergence of China as a leading nation in science.* Research policy, 2006. **35**(1): p. 83-104.
6. Cao, C., et al., *Returning scientists and the emergence of China's science system.* Science and Public Policy, 2020. **47**(2): p. 172-183.
7. Grimm, J. and I. Saliba, *Free research in fearful times: Conceptualizing an index to monitor academic freedom.* Interdisciplinary Political Studies, 2017. **3**(1): p. 41-75.
8. Wagner, C.S. and L. Leydesdorff, *Network structure, self-organization, and the growth of international collaboration in science.* Research policy, 2005. **34**(10): p. 1608-1618.
9. Bhaskar, R., *A realist theory of science*. Radical thinkers. 2008, London ; New York: Verso. 284 p.
10. Taylor, M.Z., *The politics of innovation: Why some countries are better than others at science and technology*. 2016: Oxford University Press.
11. Whetsell, T.A., et al., *Democracy, Complexity, and Science: Exploring Structural Sources of National Scientific Performance.* Science and Public Policy, 2021. **48**(5): p. 697-711.
12. Kozlowski, D., et al., *Intersectional inequalities in science.* Proc Natl Acad Sci U S A, 2022. **119**(2): p. e2113067119.
13. Waltman, L., *A review of the literature on citation impact indicators.* Journal of informetrics, 2016. **10**(2): p. 365-391.
14. Kinne, B.J., *Network dynamics and the evolution of international cooperation.* American Political Science Review, 2013. **107**(4): p. 766-785.
15. Gallop, M.B., *Endogenous networks and international cooperation.* Journal of Peace Research, 2016. **53**(3): p. 310-324.
16. Barabasi, A.L. and R. Albert, *Emergence of scaling in random networks.* Science, 1999. **286**(5439): p. 509-12.
17. Newman, M.E., *The structure and function of complex networks.* SIAM review, 2003. **45**(2): p. 167-256.
18. Barabâsi, A.-L., et al., *Evolution of the social network of scientific collaborations.* Physica A: Statistical mechanics and its applications, 2002. **311**(3-4): p. 590-614.
19. Cranmer, S.J., B.A. Desmarais, and J.W. Morgan, *Inferential network analysis*. 2020: Cambridge University Press.
20. Price, D.J.D.S. Little Science, Big Science. 1963, New York Chichester, West Sussex: Columbia University Press.
21. Castelvecchi, D., *Physics paper sets record with more than 5,000 authors.* Nature, 2015. **15**.
22. Wagner, C.S., *The new invisible college: Science for development*. 2009: Brookings Institution Press.

23. Wagner, C.S., T.A. Whetsell, and L. Leydesdorff, *Growth of international collaboration in science: revisiting six specialties.* Scientometrics, 2017. **110**(3): p. 1633-1652.
24. Wagner, C.S., H.W. Park, and L. Leydesdorff, *The Continuing Growth of Global Cooperation Networks in Research: A Conundrum for National Governments.* PLoS One, 2015. **10**(7): p. e0131816.
25. Landry, R. and N. Amara, *The impact of transaction costs on the institutional structuration of collaborative academic research.* Research policy, 1998. **27**(9): p. 901-913.
26. Huang, M.H., L.L. Wu, and Y.C. Wu, *A study of research collaboration in the pre-web and post-web stages: A coauthorship analysis of the information systems discipline.* Journal of the Association for information science and technology, 2015. **66**(4): p. 778-797.
27. Wray, K.B., *The epistemic significance of collaborative research.* Philosophy of Science, 2002. **69**(1): p. 150-168.
28. Wagner, C.S., et al., *Approaches to understanding and measuring interdisciplinary scientific research (IDR): A review of the literature.* Journal of informetrics, 2011. **5**(1): p. 14-26.
29. Katz, J.S. and B.R. Martin, *What is research collaboration?* Research policy, 1997. **26**(1): p. 1-18.
30. Defazio, D., A. Lockett, and M. Wright, *Funding incentives, collaborative dynamics and scientific productivity: Evidence from the EU framework program.* Research policy, 2009. **38**(2): p. 293-305.
31. Bozeman, B., J.S. Dietz, and M. Gaughan, *Scientific and technical human capital: an alternative model for research evaluation.* Int. J. Technol. Manag., 2001. **22**(7/8): p. 716-740.
32. Melkers, J. and A. Kiopa, *The social capital of global ties in science: The added value of international collaboration.* Review of Policy Research, 2010. **27**(4): p. 389-414.
33. Dusdal, J. and J.J. Powell, *Benefits, motivations, and challenges of international collaborative research: A sociology of science case study.* Science and Public Policy, 2021. **48**(2): p. 235-245.
34. Abramo, G., C.A. D'Angelo, and M. Solazzi, *The relationship between scientists' research performance and the degree of internationalization of their research.* Scientometrics, 2011. **86**(3): p. 629-643.
35. Fox, M.F. and I. Nikivincze, *Being highly prolific in academic science: characteristics of individuals and their departments.* Higher Education, 2021. **81**(6): p. 1237-1255.
36. Glänzel, W. and C. de Lange, *A distributional approach to multinationality measures of international scientific collaboration.* Scientometrics, 2002. **54**(1): p. 75-89.
37. Van Raan, A.F., *Science as an international enterprise.* Science and Public Policy, 1997. **24**(5): p. 290-300.
38. Sugimoto, C.R., et al., *Scientists have most impact when they're free to move.* Nature, 2017. **550**(7674): p. 29-31.
39. Chinchilla-Rodríguez, Z., et al., *Travel bans and scientific mobility: utility of asymmetry and affinity indexes to inform science policy.* Scientometrics, 2018. **116**: p. 569–590.
40. Robinson-Garcia, N., et al., *The many faces of mobility: Using bibliometric data to measure the movement of scientists.* Journal of Informetrics, 2019. **13**(1): p. 50-63.

41. Glänzel, W. and A. Schubert, *Domesticity and internationality in co-authorship, references and citations.* Scientometrics, 2005. **65**(3): p. 323-342.
42. Leydesdorff, L., L. Bornmann, and C.S. Wagner, *The Relative Influences of Government Funding and International Collaboration on Citation Impact.* J Assoc Inf Sci Technol, 2019. **70**(2): p. 198-201.
43. Subramanyam, K., *Bibliometric studies of research collaboration: A review.* Journal of information Science, 1983. **6**(1): p. 33-38.
44. Davidson Frame, J. and M.P. Carpenter, *International research collaboration.* Social studies of science, 1979. **9**(4): p. 481-497.
45. Whitley, R., *The intellectual and social organization of the sciences*. 2000: Oxford University Press on Demand.
46. Popper, K., *The open society and its enemies*. 2012: Routledge.
47. Parsons, T., *The structure of social action*. 1937, New York, N.Y.: McGraw-Hill Book Company.
48. Barber, B., *Science and the Social Order*. Vol. 30. 1962: Greenwood Press. 87-88.
49. Merton, R.K., *Science and the social order.* Philosophy of science, 1938. **5**(3): p. 321-337.
50. Merton, R.K., *The sociology of science : theoretical and empirical investigations*. 1973, Chicago: University of Chicago Press. xxxi, 605 pages.
51. Taylor, M.Z., *Toward an international relations theory of national innovation rates.* Security Studies, 2012. **21**(1): p. 113-152.
52. Wang, Q.-J., et al., *The impacts of democracy on innovation: Revisited evidence.* Technovation, 2021. **108**: p. 102333.
53. Gao, Y., et al., *Does democracy cause innovation? An empirical test of the popper hypothesis.* Research Policy, 2017. **46**(7): p. 1272-1283.
54. Wiesner, K., et al., *Stability of democracies: a complex systems perspective.* European Journal of Physics, 2018. **40**(1): p. 014002.
55. Eliassi-Rad, T., et al., *What science can do for democracy: a complexity science approach.* Humanities and Social Sciences Communications, 2020. **7**(1): p. 1-4.
56. Uzzi, B., et al., *Atypical combinations and scientific impact.* Science, 2013. **342**(6157): p. 468-72.
57. Wagner, C.S., T.A. Whetsell, and S. Mukherjee, *International research collaboration: Novelty, conventionality, and atypicality in knowledge recombination.* Research Policy, 2019. **48**(5): p. 1260-1270.
58. Dahl, R.A., *Polyarchy: Participation and opposition*. 1971: Yale university press.
59. Dahl, R.A., *Democracy and its Critics*. 1989: Yale university press.
60. Lührmann, A. and S.I. Lindberg, *A third wave of autocratization is here: what is new about it?* Democratization, 2019. **26**(7): p. 1095-1113.
61. Hellmeier, S., et al., *State of the world 2020: autocratization turns viral.* Democratization, 2021. **28**(6): p. 1053-1074.
62. Enyedi, Z., *Academic Solidarity and the Culture War in Orbán's Hungary.* PS: Political Science & Politics, 2022. **55**(3): p. 582-584.
63. Fernando, L., *The Lima declaration on academic freedom and autonomy of institutions of higher education.* Higher Education Policy, 1989. **2**(1): p. 49-51.
64. Berggren, N. and C. Bjørnskov, *Political institutions and academic freedom: evidence from across the world.* Public Choice, 2022. **190**(1): p. 205-228.

65. Newman, M.E., *Mixing patterns in networks.* Phys Rev E Stat Nonlin Soft Matter Phys, 2003. **67**(2 Pt 2): p. 026126.
66. McPherson, M., L. Smith-Lovin, and J.M. Cook, *Birds of a feather: Homophily in social networks.* Annual review of sociology, 2001: p. 415-444.
67. Mearsheimer, J.J. and G. Alterman, *The tragedy of great power politics*. 2001: WW Norton & Company.
68. Waltz, K.N., *Man, the state, and war: A theoretical analysis*. 2001, Columbia University Press.
69. Keohane, R.O., *After hegemony: Cooperation and discord in the world political economy*. 2005: Princeton university press.
70. Lai, B. and D. Reiter, *Democracy, political similarity, and international alliances, 1816-1992.* Journal of Conflict Resolution, 2000. **44**(2): p. 203-227.
71. Leeds, B.A., *Domestic political institutions, credible commitments, and international cooperation.* American Journal of Political Science, 1999: p. 979-1002.
72. Mansfield, E.D., H.V. Milner, and B.P. Rosendorff, *Why democracies cooperate more: Electoral control and international trade agreements.* International Organization, 2002. **56**(3): p. 477-513.
73. Mansfield, E.D., H.V. Milner, and B.P. Rosendorff, *Replication, realism, and robustness: Analyzing political regimes and international trade.* American Political Science Review, 2002. **96**(1): p. 167-169.
74. Kinne, B.J., *Dependent diplomacy: Signaling, strategy, and prestige in the diplomatic network.* International Studies Quarterly, 2014. **58**(2): p. 247-259.
75. Warren, T.C., *Modeling the coevolution of international and domestic institutions: Alliances, democracy, and the complex path to peace.* Journal of Peace Research, 2016. **53**(3): p. 424-441.
76. Wasserman, S. and P. Pattison, *Logit models and logistic regressions for social networks: I. An introduction to Markov graphs andp.* Psychometrika, 1996. **61**(3): p. 401-425.
77. Robins, G., et al., *An introduction to exponential random graph (p\*) models for social networks.* Social networks, 2007. **29**(2): p. 173-191.
78. Lusher, D., J. Koskinen, and G. Robins, *Exponential random graph models for social networks: Theory, methods, and applications*. 2013: Cambridge University Press.
79. Zhang, C., et al., *Understanding scientific collaboration: Homophily, transitivity, and preferential attachment.* Journal of the Association for Information Science and Technology, 2018. **69**(1): p. 72-86.
80. Abbasiharofteh, M. and T. Broekel, *Still in the shadow of the wall? The case of the Berlin biotechnology cluster.* Environment and Planning A: Economy and Space, 2021. **53**(1): p. 73-94.
81. Sciabolazza, V.L., et al., *Detecting and analyzing research communities in longitudinal scientific networks.* PloS one, 2017.
82. Akbaritabar, A., et al., *Italian sociologists: a community of disconnected groups.* Scientometrics, 2020. **124**: p. 2361–2382.
83. Akbaritabar, A. and G. Barbato, *An internationalised Europe and regionally focused Americas: A network analysis of higher education studies.* European Journal of Education, Research, Development, and Policy 2021.
84. Smith, T.B., et al., *Great minds think alike, or do they often differ? Research topic overlap and the formation of scientific teams.* Journal of Informetrics, 2021. **15**(1).

85. Hanneke, S., W. Fu, and E.P. Xing, *Discrete temporal models of social networks.* Electronic journal of statistics, 2010. **4**: p. 585-605.
86. Leifeld, P., S.J. Cranmer, and B.A. Desmarais, *Temporal exponential random graph models with btergm: Estimation and bootstrap confidence intervals.* Journal of Statistical Software, 2018. **83**: p. 1-36.
87. Cranmer, S.J. and B.A. Desmarais, *Inferential network analysis with exponential random graph models.* Political analysis, 2011. **19**(1): p. 66-86.
88. Desmarais, B.A. and S.J. Cranmer, *Statistical mechanics of networks: Estimation and uncertainty.* Physica A: Statistical Mechanics and its Applications, 2012. **391**(4): p. 1865-1876.
89. Leifeld, P. and S.J. Cranmer, *A theoretical and empirical comparison of the temporal exponential random graph model and the stochastic actor-oriented model.* Network science, 2019. **7**(1): p. 20-51.
90. Huang, P. and C.T. Butts, *Parameter Estimation Procedures for Exponential-Family Random Graph Models on Count-Valued Networks: A Comparative Simulation Study.* arXiv preprint arXiv:2111.02372, 2021.
91. Neal, Z.P., *backbone: An R package to extract network backbones.* PloS one, 2022. **17(5)**.
92. Neal, Z., *The backbone of bipartite projections: Inferring relationships from co-authorship, co-sponsorship, co-attendance and other co-behaviors.* . Social Networks, 2014. **39**: p. 84-97.
93. Krivitsky, P.N., *Exponential-family random graph models for valued networks.* Electron J Stat, 2012. **6**: p. 1100-1128.
94. Krivitsky, P.N. and C.T. Butts, *Modeling valued networks with statnet.* The Statnet Development Team, 2013: p. 2013.
95. Handcock, M.S., et al., *statnet: Software Tools for the Representation, Visualization, Analysis and Simulation of Network Data.* J Stat Softw, 2008. **24**(1): p. 1548-7660.
96. Mabry, P.L., et al., *CADRE: A Collaborative, Cloud-Based Solution for Big Bibliographic Data Research in Academic Libraries.* Front Big Data, 2020. **3**: p. 556282.
97. Stivala, A., G. Robins, and A. Lomi, *Exponential random graph model parameter estimation for very large directed networks.* PLOS ONE, 2020. **15**(1): p. e0227804.
98. Gurney, T., E. Horlings, and P. van den Besselaar, *Author disambiguation using multi-aspect similarity indicators.* Scientometrics, 2012. **91**(2): p. 435-449.
99. Balland, P.-A., et al., *Complex economic activities concentrate in large cities.* Nature Human Behavior, 2020. **4**: p. 248–254.
100. Pina, D.G., et al., *Effects of seniority, gender and geography on the bibliometric output and collaboration networks of European Research Council (ERC) grant recipients.* PloS one, 2019.
101. Akbaritabar, A., *A quantitative view of the structure of institutional scientific collaborations using the example of Berlin.* Quantitative Science Studies, 2021. **2**(2): p. 753–777.
102. Pilny, A. and Y. Atouba, *Modeling valued organizational communication networks using exponential random graph models.* Management Communication Quarterly, 2018. **32**(2): p. 250-264.
103. Fruchterman, T.M. and E.M. Reingold, *Graph drawing by force-directed placement.* Software: Practice and experience, 1991. **21**(11): p. 1129-1164.

104. Tyner, S.C., F. Briatte, and H. Hofmann, *Network Visualization with ggplot2.* The R Journal, 2017.
105. Duxbury, S.W., *Diagnosing multicollinearity in exponential random graph models.* Sociological Methods & Research, 2021. **50**(2): p. 491-530.